\documentclass{aa}  

\usepackage{float}
\usepackage{epstopdf}
\usepackage{booktabs}
\usepackage{multirow}
\usepackage{textcomp}
\usepackage{graphicx}
 \usepackage{color}
\usepackage{array}
\newcolumntype{H}{>{\setbox0=\hbox\bgroup}c<{\egroup}@{}}
\usepackage{txfonts}
\begin{document}

   \title{Radio halos in a mass-selected sample of 75 galaxy clusters}
   \subtitle{Paper II - Statistical analysis}

\authorrunning{V. Cuciti et al.}
  \author{V. Cuciti
          \inst{1},
          R. Cassano\inst{2},
          G. Brunetti \inst{2}, D. Dallacasa\inst{3,2},
          F. de Gasperin\inst{1},
          S. Ettori\inst{7,8},
          S. Giacintucci\inst{9},
          R. Kale\inst{5},
          G.~W. Pratt\inst{6},
          R.~J. van Weeren\inst{4},
          T. Venturi\inst{2}
          }

\institute{Hamburger Sternwarte, Universit\"at Hamburg, Gojenbergsweg 112, 21029, Hamburg, Germany,
\email{vcuciti@hs.uni-hamburg.de} \and INAF-Istituto di Radioastronomia, via P. Gobetti 101, 40129 Bologna, Italy
\and Dipartimento di Fisica e Astronomia, Universit\`{a} di Bologna, via P. Gobetti 93/2, 40129 Bologna, Italy
\and Leiden Observatory, Leiden University, PO Box 9513, 2300 RA Leiden, The Netherlands
\and National Centre for Radio Astrophysics, Tata Institute of Fundamental Research       Savitribai Phule Pune University Campus, Pune 411 007 Maharashtra, INDIA
\and AIM, CEA, CNRS, Université Paris-Saclay, Université Paris Diderot, Sorbonne Paris Cité, F-91191 Gif-sur-Yvette, France
\and INAF, Osservatorio di Astrofisica e Scienza dello Spazio, via Pietro Gobetti 93/3, 40129 Bologna, Italy 
\and
INFN, Sezione di Bologna, viale Berti Pichat 6/2, I-40127 Bologna
\and Naval Research Laboratory, 4555 Overlook Avenue SW, Code 7213, Washington, DC 20375, USA}

   \date{Received --; accepted --}

 
  \abstract
   {Many galaxy clusters host Mpc scale diffuse radio sources called radio halos. Their origin is tightly connected to the processes that lead to the formation of clusters themselves. In order to unveil this connection, statistical studies of the radio properties of clusters combined with their thermal properties are necessary. To this purpose, we selected a sample of galaxy clusters with $M_{500}\geq6\times10^{14} \,M_\odot$ and $z=0.08-0.33$ from the Planck SZ catalogue. In paper I, we presented the radio and X-ray data analysis that we carried out on the clusters of this sample. }
   {In this paper, we exploited the wealth of data presented in paper I to study the radio properties of the sample, in connection to the mass and dynamical state of clusters. }
   {We used the dynamical information derived from the X-ray data to assess the role of mergers in the origin of radio halos. We studied the distribution of clusters in the radio power -- mass diagram, the scaling between the radio luminosity of radio halos and the mass of the host clusters and the role of dynamics on the radio luminosity and emissivity of radio halos. We measured the occurrence of radio halos as a function of the cluster mass and we compared it with the expectations of models developed in the framework of turbulent acceleration.}
   {We found that more than the 90\% of radio halos are in merging clusters and that their radio power correlates with the mass of the host clusters. The correlation shows a large dispersion. Interestingly, we showed that cluster dynamics contributes significantly to this dispersion with more disturbed clusters being more radio luminous. Clusters without radio halos are generally relaxed and the upper limits to their diffuse emission lie below the correlation. Moreover, we showed that the radio emissivity of clusters exhibits an apparent bimodality, with the emissivity of radio halos being at least $\sim5$ times larger than the non-emission associated with more relaxed clusters. We found that the fraction of radio halos drops from $\sim70\%$ in high mass clusters to $\sim35\%$ in the lower mass systems of the sample and we showed that this result is in good agreement with the expectations from turbulent re-acceleration models.}
   {}

   \maketitle
%

\section{Introduction}
An increasing number of galaxy clusters show diffuse synchrotron emission detected in the radio band. This emission reveals that the Intra Cluster Medium (ICM) is permeated with relativistic particles and magnetic fields. Cluster-scale radio sources can be in form of radio halos, radio relics or mini halos \citep[][for a review]{vanweeren19}. Radio halos are the main focus of this paper. They are found at the centre of some merging galaxy clusters and have typical size of $1-2$ Mpc. 

In current models, radio halos have origin when electrons in the ICM are re-accelerated by turbulence, injected during merger events \citep[][for a review]{brunettijones14}. In this scenario, the properties of radio halos should be connected to the mass and to the merging history of the host clusters. The statistical study of large samples of galaxy clusters with deep radio observations is necessary to investigate this connection. In this respect, a big step forward has been done during the past few decades, mainly thanks to the combination of radio and X-ray observations \citep[e.g.][]{giovannini99,kempner01, liang00,venturi07,venturi08,kale13,kale15}. It has been shown that the radio power of radio halos correlates with the X-ray luminosity of the host clusters \citep[e.g.][]{liang00, brunetti09}. Moreover, clusters without radio halos (for which upper limits are available) lie well below this correlation \citep{brunetti07}. While the connection between the presence of radio halos and the clusters' merging state had been previously proposed for a few cases \citep{buote01, markevitch01, govoni04}, \citet{cassano10} found the first statistical evidence that radio halos are predominately found in merging clusters, while clusters without halo detection are generally relaxed. More recently, similar studies have been also performed in samples selected through Sunyaev-Zel'dovich (SZ) measurements. This was a major improvement in the study of diffuse emission in clusters, because the SZ effect is currently the best available proxy of the cluster mass \citep[e.g.][]{nagai06}, which is the key parameter in the models for the formation of radio halos, as it sets the energy budget available for particle acceleration. A correlation has been found also between the synchrotron power of radio halos and the mass of the host clusters \citep{basu12,cassano13}. Although earlier studies were unable to find a clear segregation between radio halos and upper limits in SZ-selected clusters \citep{basu12}, thanks to the improved statistics, \citet{cassano13} showed that a bimodal behaviour is also observed in the radio power-mass diagram. \citet{sommerbasu14} showed that the fraction of radio halos in mass-selected sample is larger with respect to X-ray selected samples.

To perform a solid statistical study of radio halos in a large mass-selected sample of galaxy clusters, we selected 75 clusters with $M_{500}\footnote{$M_{500}$ is the mass enclosed in a sphere with radius $R_{500}$, defined as the radius within which the mean
mass over-density of the cluster is 500 times the cosmic critical density at the cluster redshift}\geq6\times10^{14} \,M_\odot$ and $z=0.08-0.33$ from the Planck SZ catalogue \citep{planck14}. In \cite{cuciti15}, we analysed the occurrence of radio halos in a subsample of 57 clusters with available literature information mainly coming from previous works based on X-ray selected samples \citep[e.g.][]{venturi07, venturi08}. We showed that the fraction of radio halos drops in low mass systems. However, that result, could be affected by the incompleteness of the sample, and we therefore carried out a radio observational campaign to cover all the remaining clusters. Thanks to these observations, this is now the largest ($>80\%$ complete) mass-selected sample of clusters with complete deep radio observations {\footnote The typical sensitivity of these observations is $\sim 50-100$ mJy/beam at 610 MHz and $\sim 20-70$ mJy/beam at 1.4 GHz, see paper I}. In paper I, we presented the results of these new observations and we summarised the radio and X-ray properties of the sample drawn from the observations. Here, we exploit the wealth of information acquired on the sample to
perform the statistical analysis of the radio properties of these clusters, in connection to their mass and dynamics. In particular, now that the limitations associated to the incompleteness of the subsample in \cite{cuciti15} have been overcome, we aim to address: the distribution of galaxy clusters in the radio power-mass diagram, the difference in 
`radio loudness' between merging and non merging clusters, the occurrence of radio halos as a function of the cluster mass and its comparison with model expectations.

In Section \ref{sec:sample} we briefly summarise the radio properties of the sample. In Section \ref{Sec:connection} we discuss the connection between the presence of radio halos and the disturbed dynamical state of clusters.
We analyse the distribution of clusters in the radio power--mass and radio emissivity--mass diagrams in Section \ref{Sec:PM}. We measure the occurrence of radio halos as a function of the cluster mass in Section \ref{Sec:occurrence} and we compare it with the model expectations. In Section \ref{Sec:conclusion}, we summarise the main results of this work. Throughout this paper we assume a $\Lambda$CDM cosmology with $H_0=70$ km s$^{-1}$Mpc$^{-1}$, $\Omega_\Lambda=0.7$ and $\Omega_m=0.3$.

\section{The sample}   
\label{sec:sample}
The sample analysed in this paper is composed of 75 galaxy clusters, selected from the Planck SZ cluster catalogue. The selection criteria, extensively discussed in \citet{cuciti15} and paper I, can be summarised as follows:
\begin{itemize}
\item at redshift $0.08<z<0.2$ we selected clusters with $M_{500}\geq5.7\times10^{14}M_\odot$
\item at redshift $0.2<z<0.33$ we selected clusters with $M_{500}\geq6\times10^{14} \,M_\odot$
\end{itemize}

Combining literature information with our new observations, all these clusters have available radio information at $\sim$ GHz frequencies. 63 clusters also have archival Chandra X-ray observations. The radio and X-ray data analysis are discussed in paper I. Here, we summarise the diffuse sources present in our sample:
\begin{itemize}
\item 28 ($\sim$37\%) clusters host radio halos, 10 of them are USSRH (Ultra Steep Spectrum) or candidate USSRH
\item seven ($\sim$10\%) clusters have radio relics, five of them also have radio halos (and have been already counted above)
\item 11 ($\sim$15\%) clusters host mini halos 
\end{itemize}

Moreover, we found candidate diffuse emission in six clusters, one is a candidate mini halo and five are candidate radio halos. 31
($\sim$41\%) clusters do not show any hint of central diffuse emission at the sensitivity of current observations. Combining the work done in paper I with literature information, we were able to derive upper limits for 22 of these clusters. For clusters with GMRT data only ($\sim 50\%$), we scaled the upper limits to 1.4 GHz assuming a typical spectral index $\alpha=-1.3$ \citep{vanweeren19}. The lack of upper limits for the remaining 9 clusters is mainly due to the bad data quality or to the presence of artifacts around bright sources affecting the cluster's field. We note that these 9 clusters do not show any significant deviation in terms of dynamics with respect to the 22 clusters with available upper limits. Moreover, they are uniformly distributed in mass. For these reasons we do not expect the lack of these upper limits to significantly affect the results presented in this paper.

\section{Radio halo--merger connection}
\label{Sec:connection}

\begin{figure*}
\centering
\includegraphics[scale=0.4]{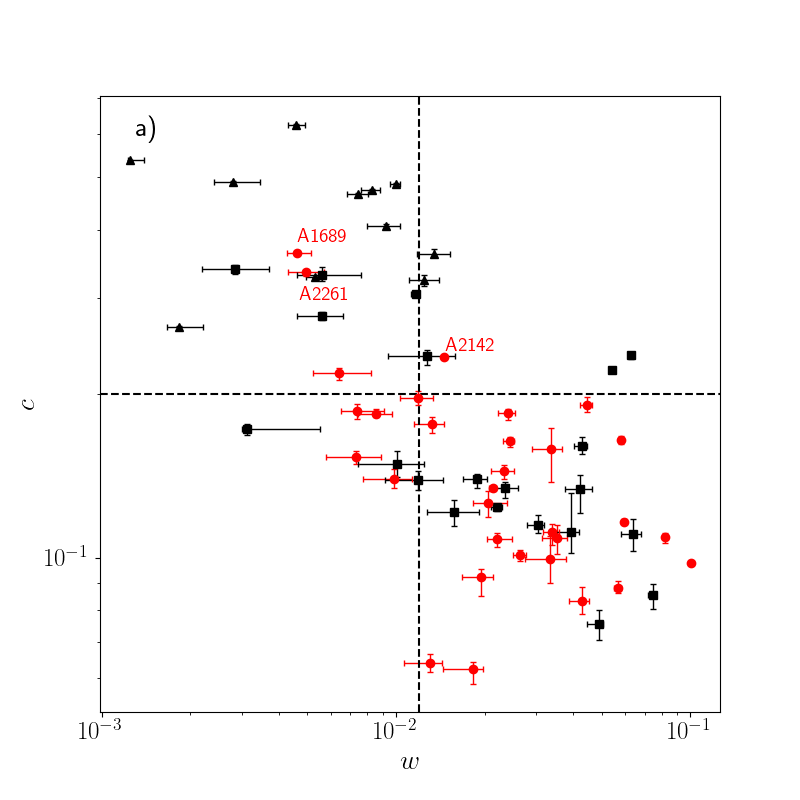}
\includegraphics[scale=0.4]{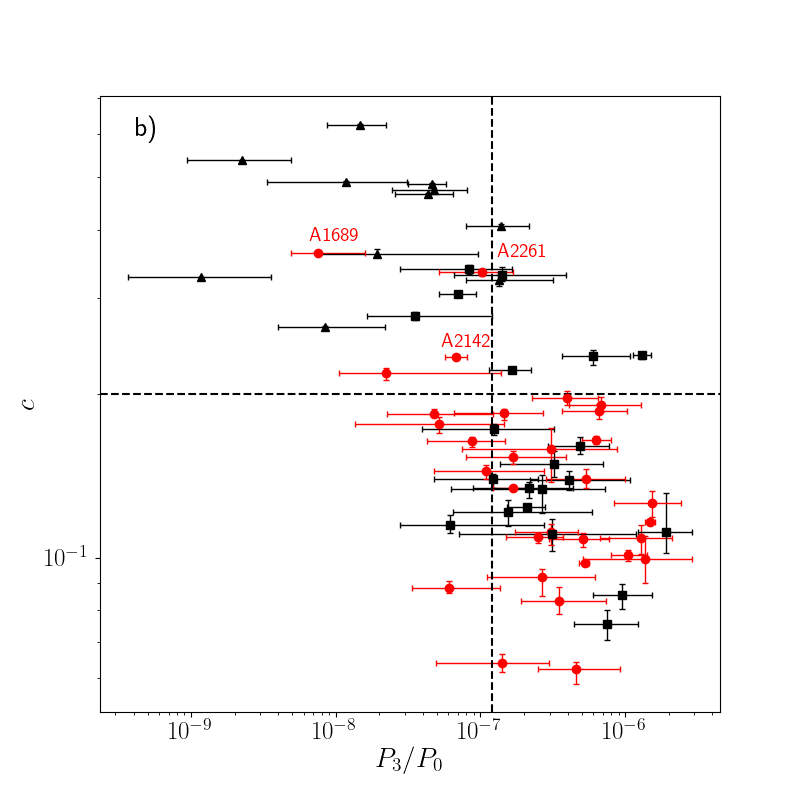}
\includegraphics[scale=0.4]{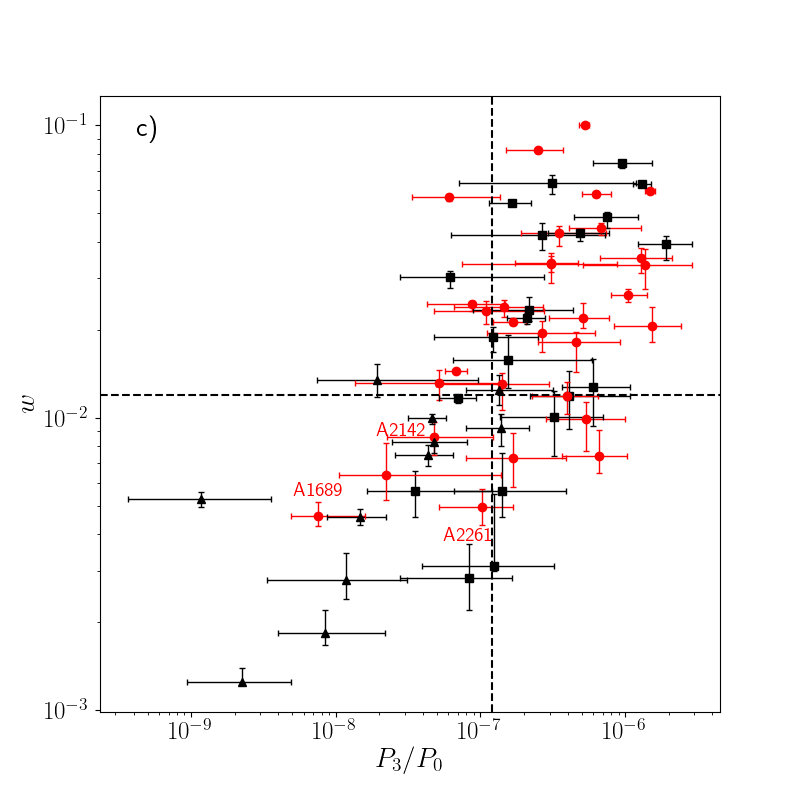}
\caption{a) $c-w$, b) $c-P_3/P_0$, c) $w-P_3/P_0$ morphological diagrams for the clusters in our sample with available X-ray \textit{Chandra} data. Vertical and horizontal dashed lines are adapted from \citet{cassano10} and are: $c=0.2$, $w=0.012$, and $P_{3}/P_{0}=1.2\times10^{-7}$. Red circles are radio halos. Black squares represent clusters without radio halos. Black triangles are clusters with mini halos. The values of the parameters plotted here are listed in paper I, Table 5. The names label clusters that are explicitly mentioned in the text.}
\label{Fig:morphological_diagrams}
\end{figure*}

It is widely accepted that radio halos are preferentially found in merging clusters \citep[e.g.][]{cassano10,cassano13,bonafede15, kale15,cuciti15}. In this context, our sample offers the opportunity to assess the radio halo merger connection in a large and mass-complete sample of clusters. 
We used three morphological parameters that are largely used in the literature to define the dynamical state of clusters: the concentration parameter, $c$, the centroid shift, $w$ and the power ratios, $P_3/P_0$ \citep[][see paper I]{buote01, santos08, lovisari17, andrade-santos17, rossetti17}. Fig. \ref{Fig:morphological_diagrams} shows the three morphological diagrams $c-w$, $c-P_3/P_0$ and $w-P_3/P_0$. 
The dashed lines are adapted from \citet{cassano10} and correspond to $c=0.2, w= 0.012$ and $P_3/P_0=1.2\times10^{-7}$. In these plots, relaxed clusters are in the regions with high values of $c$ and low values of $w$ and $P_3/P_0$, and they become gradually more disturbed going towards the opposite corner of the diagrams.
Mini halos are marked as triangles in Fig. \ref{Fig:morphological_diagrams} and they all lie in the regions of relaxed systems. According to the morphological parameters classification and considering that A1689 (that would be classified as relaxed based on morphological parameters) is a merging system with the merger occurring along the line of sight \citep{andersson04}, more than the 90\% of radio halos in this sample are hosted by merging clusters, while only two of them ($<$10\%) are found in more relaxed systems. These two clusters are A2142 and A2261. A2142 is a minor merging systems hosting multiple Mpc-scale cold fronts \citep{markevitch00, owers11, rossetti13, venturi17} and A2261 would be classified as a dynamically intermediate cluster from our visual inspection (paper I), due to the presence of substructures on large scale. 

\begin{figure*}
\centering
\includegraphics[scale=0.45]{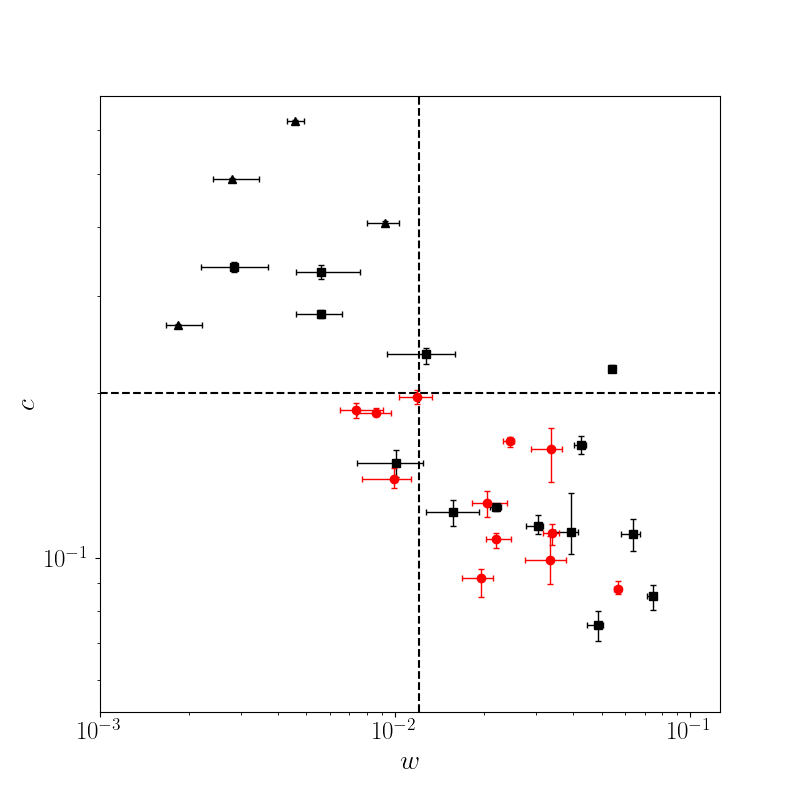}
\includegraphics[scale=0.45]{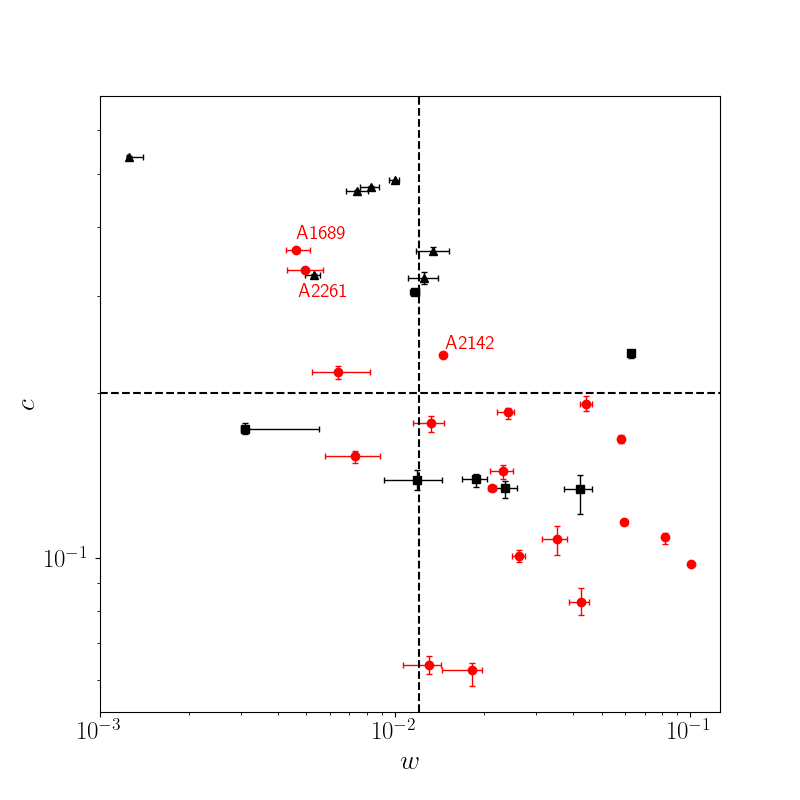}
\caption{$c-w$ diagram for low mass (\textit{left}) and high mass (\textit{right)} clusters. Symbols are the same as in Fig. \ref{Fig:morphological_diagrams}.}
\label{Fig:c_w_massa}
\end{figure*}

Fig.\ref{Fig:morphological_diagrams} shows that only $\sim60$\% of merging clusters host radio halos, confirming previous studies \citep[e.g.][]{cassano13,bonafede17}. This suggests that mergers are not the unique players in the formation of radio halos \citep{brunettijones14}. One major player is the mass of the clusters that sets the amount of energy released during merger events. To investigate the role of the cluster mass in the radio halo merger connection, we split the clusters into two sub-samples according to the median mass value of $M_{500}=7\times10^{14}M_\odot$, which guarantees equal statistics in each sub-sample.
We focus on $c$ and $w$ because they are the most robust parameters to define the dynamical state of clusters and they are also determined with much higher accuracy with respect to the power ratios (paper I and references therein). Fig. \ref{Fig:c_w_massa} shows that the fraction of merging clusters without radio halos is $\sim$20\% in high mass systems, while it is $\sim$65\% in low mass objects.
If we attempt to interpret this evidence in the context of turbulent re-acceleration models, it is the consequence of the fact that massive and merging systems form radio halos emitting up to high frequency, while merging events in low mass cluster may not induce enough turbulence to accelerate particles up to the energies necessary to emit radiation at GHz frequency \citep[][]{CBS06,cassano10}. In this scenario, a large fraction of these lower mass merging clusters should host USSRH \citep[e.g.][]{brunetti08nature, cassano12}. 

Fig. \ref{Fig:c_w_massa} also shows that the two clusters with radio halos that appear dynamically relaxed from the morphological parameters (A2142 and A2261) are both in the high-mass bin. This suggests that, in massive clusters, a minor merger may be sufficient to generate radio halos even though the X-ray morphology of the cluster does not look extremely disturbed, whilst low mass-systems need major mergers introducing a larger amount of turbulent energy in the ICM to accelerate electrons and produce radio diffuse emission.

\section{Radio power--mass diagram}
\label{Sec:PM}

In this Section we investigate the distribution of clusters, with and without radio halos, in the radio power--mass diagram. The values of $P_{1.4 \rm{GHz}}$ and $M_{500}$ are listed in Table 1 of paper I. 

All radio halos of the sample, with the exception of three USSRHs, namely A1132 \citep{wilber18}, RXCJ1514.9-1523 \citep{giacintucci11}, and RXCJ1314.4-2515 \citep{venturi07}, have radio powers measured at 1.4 GHz. For these three USSRHs, that have been measured only at low radio frequency, we extrapolated their 1.4 GHz radio powers with the estimated spectral index and we adopted a conservative uncertainty of 30--50\%, corresponding to a variation in $\alpha$ of 0.3-0.4.

\subsection{Fitting procedure}
\label{sec:fitting_procedure}
We followed the fitting procedure outlined in \citet{cassano13}. Specifically, we fit a power-law relation in the log--log space by adopting the BCES linear regression algorithms \citep{akritas96} which take measurement errors in both variables into account.
We fitted the observed $P_{1.4 \rm{GHz}}-M_{500}$ data points with a power-law in the form:
\begin{equation}
\mathrm{log}\left(\frac{P_{1.4 \rm{GHz}}}{10^{24.5}\mathrm{W/Hz}}\right)=B~\mathrm{log}\left(\frac{M_{500}}{10^{14.9}\,M_\odot}\right)+A\label{eq:Pot-M1}
\end{equation}
where $A$ and $B$ are the intercept and the slope of the correlation, respectively.

Considering $Y=\log(P_{1.4 \rm{GHz}})-24.5$ and $X=\log(M_{500})-14.9$, and having a sample of $N$ data points ($X_i,Y_i)$ with errors $(\sigma_{X_i},\sigma_{Y_i})$ the raw scatter of the correlation can be estimated as:
\begin{equation}
\sigma_{raw}^2=\dfrac{1}{N-2}\sum_{i=0}^N w_i (Y_i-BX_i-A)^2
\end{equation}
where
\begin{equation}
w_i=\dfrac{1/\sigma_i^2}{(1/N)\sum_{i=0}^N1/\sigma_i^2}~~~~\mathrm{and}~~~~\sigma_i^2=\sigma_{Y_i}^2+B^2\sigma_{X_i}^2
\end{equation}

Since we are dealing with a limited sample, we obtain a sampled regression line that can deviate from the true (unknown) regression line. To evaluate the 95\% confidence region of the best-fit relation, that is to say the area that has the 95\% probability of containing the `true' regression line, we calculated the 95\% confidence interval of the mean value of $Y$, $\left\langle Y\right\rangle $. For a given $X$, this is $\left\langle Y\right\rangle\pm\Delta Y $, where:
\begin{equation}
\Delta Y=\pm1.96 \sqrt{\bigg[\sum_{i=0}^N\dfrac{(Y_i-Y_m)^2}{N-2}\bigg]\bigg[\dfrac{1}{N}+\dfrac{(X-X_m)^2}{\sum_{i=0}^N(X_i-X_m)^2}\bigg]}
\end{equation}
where $Y_m=BX_i+A$ and $X_m=\sum_{i=0}^NX_i/N$ for each observed $X_i$.

\subsection{Results of the fitting}
\label{sec:fitting_results}

\begin{figure*}
\centering
\includegraphics[height=9cm]{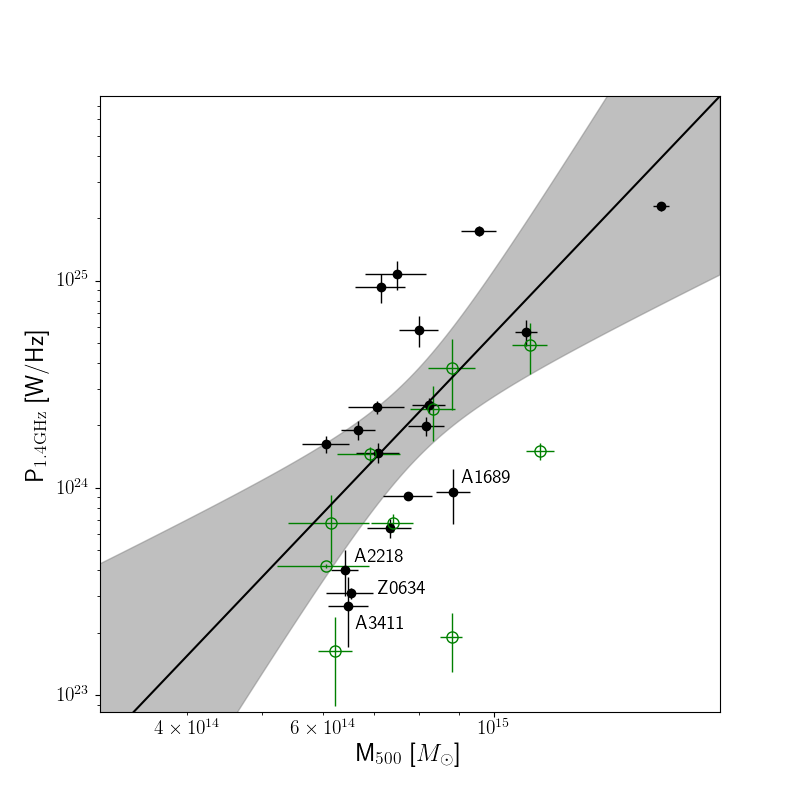}
\includegraphics[height=9cm]{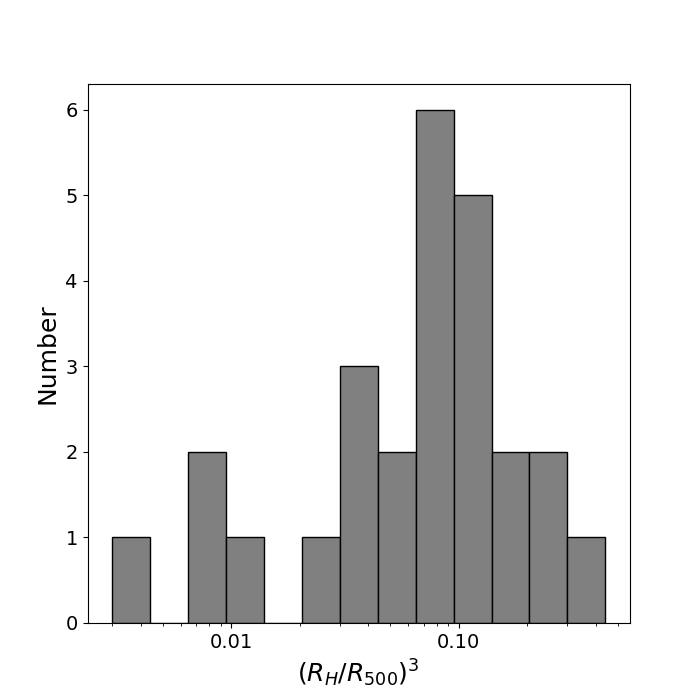}
\caption{{Radio power-mass diagram and volume distribution of the radio halos of the statistical sample. \textit Left}: $P_{1.4 \rm{GHz}}-M_{500}$ diagram. Black filled circles represent radio halos, green empty circles represent USSRHs and candidate USSRHs. The small radio halos are labelled. The best fit relation obtained with the BCES Y|X method and excluding USSRHs is shown with its 95\% confidence region. The best fit parameters are B=$3.92\pm0.79$ and A=$-0.15\pm0.10$. {\textit Right}: Distribution of the volumes of the radio halos, scaled for $R_{500}$. The four clusters lying on the left tail, separated from the main distribution are A3411, Z0634, A2218 and A1689.}
\label{Fig:histo_rh}
\end{figure*}

Different fitting methods may give different results \citep[e.g.][]{isobe90}. Therefore, it is important to choose the most suited regression method depending on the data in hand. Based on previous studies \citep[e.g.][]{cassano13, martinezaviles16}, we expect the correlation between the radio power of radio halos and the mass of clusters to be steep and the radio powers to show a large scatter around the correlation. The large scatter is due to the superposition of clusters with different merging histories, in different phases of the merger events and with different radio spectra \citep[e.g.][]{donnert13}.
Thus, it is reasonable to treat the radio power as the dependent variable and use the Y|X fitting method, which minimises the residuals in the Y variable. Another possibility, is to assume that both variables are quasi-independent and treat them symmetrically, using either the bisector method, which represents the bisector between the Y|X and X|Y regression lines, or the orthogonal method, which minimises the orthogonal distances. All these methods are available among the BCES linear regression algorithms \citep{akritas96}. In the following, we will report the results obtained with the Y|X and bisector methods.

We show the distribution of radio halos and USSRHs of our sample (which we will refer to as the `statistical' sample) in the radio power--mass diagram in the left panel of Fig. \ref{Fig:histo_rh}, together with the best fit line for radio halos only obtained with the BCES Y|X method (see Figure caption). 
In line with previous findings \citep[e.g.][]{cassano13}, USSRHs and candidate USSRHs are on or below the correlation. Even considering only radio halos with classical spectrum, we note a significant scatter around the correlation. A possible source of the scatter is a different emitting volume of the halos. For example, in \citet{cuciti18} we found two small radio halos lying below the correlation. 
In these cases, it is reasonable to expect that they have a synchrotron emissivity similar to that of larger radio halos, but they are less luminous simply because the total luminosity is produced in a smaller volume. To investigate this, we measured the radii of the radio halos as $R_H=\sqrt{R_{min}\times R_{max}}$, where $R_{min}$ and $R_{max}$ are the maximum and minimum radii measured on the 3$\sigma$ isophote, respectively \citep{cassano07}\footnote{For elongated radio halos ($R_{max}>>R_{min}$) we multiplied $R_H$ for a factor $\sqrt{\frac{R_{min}}{R_{max}}}$ to take the ellipticity of radio halos into account.}. We show the distribution of $(R_H/R_{500})^3$, which is proportional to the volume of radio halos, in the right panel of Fig. \ref{Fig:histo_rh}. The majority of radio halos are distributed around $(R_H/R_{500})^3=0.06-0.15$.

There are four radio halos whose volumes are $\sim 6-20$ times smaller than the average radio halos in our sample. They are in the clusters Z0634, A3411, A1689 and A2218 and they are all underluminous with respect to the correlation (Fig. \ref{Fig:histo_rh}, left). Interestingly, the size of these four small radio halos roughly coincides with the threshold value between radio halos and mini halos ($R_H\approx0.2\times R_{500}$) defined by \citet{giacintucci17}. On the other hand, no halos with volume larger than $5-6$ times that of the average halos are found.

Since the radio luminosity of these four radio halos is generated in a much smaller volume, their radio power cannot be directly compared to the power of giant ($R_H\sim$ Mpc) radio halos\footnote{An additional possibility is that these radio halos appear small because our observations did not recover their entire extension. However, this would also imply that a large fraction of their flux is lost by our observations and that their luminosities are biased low compared to those of the other halos in our sample.}. Also, their radio power cannot be directly compared to the upper limits because upper limits are derived using $R_H=500$ kpc by choice (corresponding to about $R_H/R_{500} \sim 0.4$ for the typical masses and redshift in our sample). Indeed, if we inject the fake radio halos on smaller scales, the upper limits would be deeper \citep{brunetti07}. Therefore, we removed them from the radio luminosity mass diagram and we performed the fit again.

\begin{table} 
\begin{footnotesize}
\begin{center}
\caption{Fitting parameters}
\begin{tabular}{lcccccc}
\hline
\hline		
Method	&	B	&	err B	&	A	&	err A &$\sigma_{raw}$ & r$_s$\\
\hline\\
\multicolumn{7}{c}{Statistical sample}\\
\hline
\textbf{RH only}		&	&	&	&  &  &\\
BCES Y|X    	& 2.96	& 0.50	& 0.01	& 0.10& 0.32 & 0.56\\
Bootstrap		& 3.32	& 3.61	& 0.02	& 0.10& &\\
BCES bisector	& 4.18	& 0.20	& $-$0.002	& 0.11 & 0.41 &\\
Bootstrap		& 4.73	& 2.61	& $-$0.02	& 0.13& & \\
\textbf{RH+USSRH}		&	&	&	& & &\\
BCES Y|X     			& 3.26	& 0.74	& $-$0.21	& 0.09& 0.41 &\\
Bootstrap				& 3.30	& 1.06	& $-$0.21	& 0.1& &\\
BCES bisector			& 4.90	& 0.16	& $-$0.22	& 0.10 & 0.52 &\\
Bootstrap				& 4.85	& 1.28	& $-$0.21	& 0.10& &\\
\hline\\		
\multicolumn{7}{c}{Extended sample}\\
\hline
\textbf{RH only}		&	&	&	& & &\\
BCES Y|X        		& 2.67	& 0.35	& 0.07	& 0.07& 0.31 & 0.64\\
Bootstrap				& 2.69	& 0.51	& 0.07	& 0.09& &\\
BCES bisector			& 3.49	& 0.14	& 0.09	& 0.08& 0.34 & \\
Bootstrap				& 3.55	& 0.63	& 0.09	& 0.09& &\\
\textbf{RH+USSRH}		&	&	&	& & &\\
BCES Y|X     			& 2.66	& 0.57	& $-$0.13	& 0.08& 0.40 &\\
Bootstrap				& 2.66	& 0.67	& $-$0.13	& 0.09& &\\
BCES bisector			& 3.97 	& 0.14 	& $-$0.11	& 0.08& 0.44 &\\
Bootstrap				& 3.91	& 0.72	& $-$0.11	& 0.09& &\\

\hline\\

\textbf{RH only}		&	&	&	& & &\\
EM algorithm			& 2.52	& 0.57	& 0.07	& 0.08& &\\
\textbf{RH+UL}			&	&	&	& & &\\
EM algorithm			& 4.90	& 1.13	& -0.52	& 0.15& &\\
\hline
\hline
\end{tabular}	
\label{Tab:fit}	
\end{center}
\end{footnotesize}
\end{table}

\begin{figure*}
\centering
\includegraphics[height=9cm]{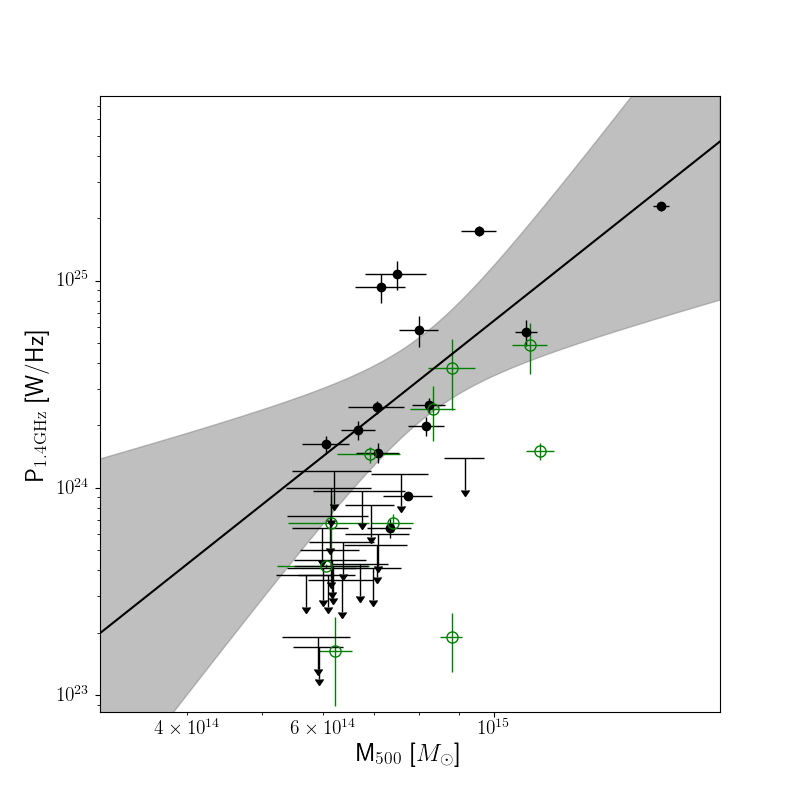}
\includegraphics[height=9cm]{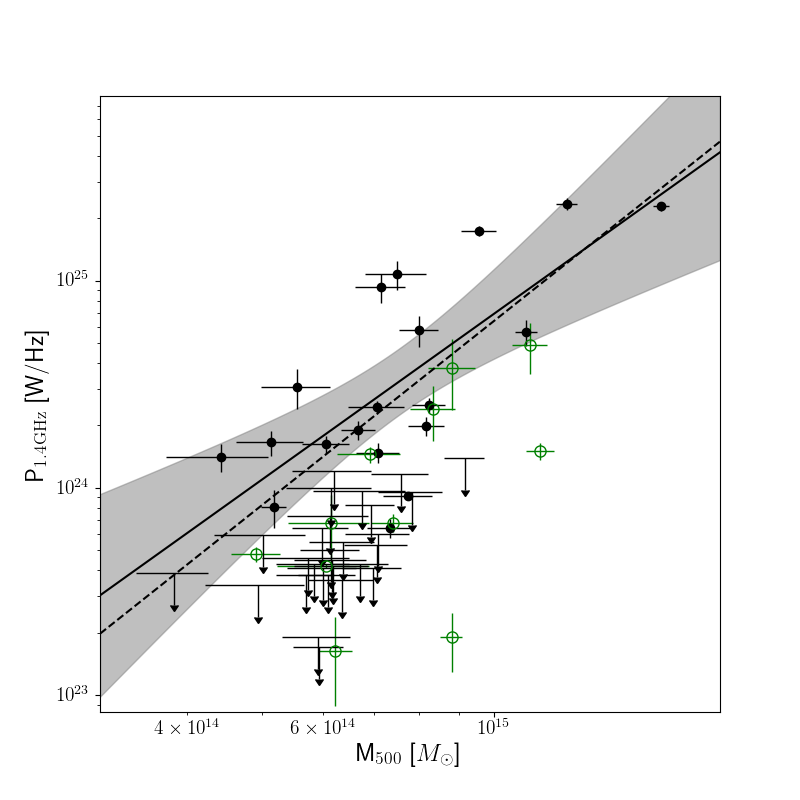}
\caption{$P_{1.4 \rm{GHz}}-M_{500}$ diagram for the clusters of the statistical sample (left) and extended sample (right). Black filled circles are radio halos, arrows are upper limits and green empty circles are USSRH or candidate USSRH. The black line and grey shadowed region show the best-fit relations (using the BCES Y|X method) and 95\% confidence region to radio halos only. For a clear comparison, in the right panel, we also show the best fit relation obtained for the statistical sample (left panel) as a dashed line.}
\label{Fig:P_M_Cuciti}
\end{figure*}

The correlation for the statistical sample without small radio halos is shown in the left panel of Fig. \ref{Fig:P_M_Cuciti} and in the top panel of Table \ref{Tab:fit} we report the best fit parameters obtained with the BCES Y|X and bisector methods, together with those from a 5000 bootstrap resampling analysis \citep[see][]{akritas96}. We report also the raw scatter of the correlation and the Spearman coefficient, $r_s$, which is a measure of the monotonicity of the relationship between two variables \footnote{$r_s=-1$ implies a perfect monotonically decreasing relation, $r_s=+1$ implies a perfect monotonically increasing relation and $r_s=0$ implies no correlation.} \citep{spearman}. We note that the bootstrap confidence intervals are quite large meaning that the fitting is likely undetermined. This can be a consequence of the cut in mass at $M_{500}\geq6\times 10^{14} M_\odot$ of our sample, that implies a small mass range ($<0.4$ in log space) to estimate the correlation coefficients. 

To evaluate the possible effect of this cut on the regression analysis, we performed Montecarlo simulations that we describe in Appendix \ref{app:bces}. We randomly distributed a number of clusters with masses in the range $M_{500}=(1-16)\times 10^{14} M_\odot$ on a known correlation (`true' correlation, that we assume to extend as a power law also at low masses) and we applied a fixed scatter in both Y and X to distribute them around the correlation. Then, we selected from the distribution only clusters with $M_{500}\geq6\times 10^{14} M_\odot$. We repeated these two steps 500 times and each time we performed the linear regression with the three methods mentioned above, comparing the resulting slopes with the `true' one. We found that, in the presence of a steep correlation with a fairly large scatter, all the fitting methods tend to give steeper slopes once the cut in the X variable is introduced. In particular, the orthogonal method is the most affected by the presence of the cut, giving the most significant deviations from the true slopes. 
According to our Montecarlo simulations we expect the `true' correlation to be $\sim 0.2-0.3$ flatter with respect to what we find in our sample if we focus on the Y|X method and we assume the scatter to be predominantly in the Y axis (see Fig. \ref{Fig:test_bces}, bottom left).


\begin{table}
\renewcommand{\arraystretch}{1.5}
\begin{center}
\begin{small}
\caption{Properties of the added clusters}
\begin{tabular}{lccc}
Name         &	z&         $M_{500}$         &        $P_{1.4 \rm{GHz}}$     \\
       &      & ($10^{14}$ M$_{\odot}$)  & ($10^{24}$ W Hz$^{-1}$)     \\
\hline

\hline\\
A545   &	 0.154	&    $4.43_{0.66}^{0.62}$    &       $1.411\pm0.22$   \\ 
Bullet &	 0.296	&    $12.41_{0.40}^{0.40}$   &       $23.44\pm1.51$   \\ 
A2255  & 0.081	&    $5.19_{0.19}^{0.19}$    &       $0.81 \pm0.17$ \\ 
A1995  & 0.319	&    $5.15_{0.52}^{0.49}$    &       $1.66\pm0.23$  \\ 
\hline\\
A746&	0.232	&    $5.56_{0.57}^{0.53}$     &       $3.07\pm0.68$  	     \\ 
A2034$^{US}$	&	0.113	&	$5.85_{0.36}^{0.35}$ 	&		$0.48\pm0.04$	\\
\hline\\
A2645		&	0.251	&	$5.02_{0.67}^{0.62}$ 		&	0.59	\\	
A267			&   0.230	&	$4.95_{0.72}^{0.67}$   	&	0.34		\\
RXJ0439.0+0715	&	0.244	&	$5.75_{0.71}^{0.70}$ 	&	0.46		\\
A611		&	0.288	&	$5.85_{0.64}^{0.60}$ 	&	0.43		\\
A2146	&	0.234	&	$3.85_{0.41}^{0.39}$ 	&	0.39		\\
\hline
\hline
\end{tabular}
\begin{flushleft}
Notes -- Top panel: radio halos from \citet{cassano13}; Middle panel: radio halos from \citet{martinezaviles16}; Bottom panel: upper limits from \citet{cassano13}.
\label{Tab:additional}
\end{flushleft}
\end{small}
\end{center}		         
\end{table}

We attempted to mitigate the limitation due to the small range in mass spanned by the clusters of the statistical sample by considering also an `extended' sample, resulting from the combination with a sub-sample from \citet{cassano13} and \citet{martinezaviles16}. In particular, we added only clusters in the same redshift range of our sample (six radio halos). Since one of the goals of this Section is to study the distribution of radio halos and upper limits in the radio power--mass diagram (see Section \ref{Sec:bimodality}), we added also the upper limits (five) from \citet{cassano13} in the same redshift range of our sample.
The properties of these additional clusters are listed in Table \ref{Tab:additional}. We point out that all these six radio halos have radio powers measured at 1.4 GHz. This allows us to study the scaling relation with larger statistics and, most importantly, within a larger mass range.  

The results obtained by applying the fitting procedure to the radio halos of the extended sample are summarised in the bottom panel of Table \ref{Tab:fit} and shown in the right panel of Fig. \ref{Fig:P_M_Cuciti}. We note that the bootstrap confidence intervals are now more in agreement with the analytic estimates, meaning that the fitting parameters are now better determined.

Although they agree with each other within the 1-$\sigma$ uncertainties, we note that the correlation derived for the extended sample, is slightly flatter, of $\sim0.3$, compared to that derived for the statistical sample. This is in line with the outcome of our Montecarlo simulations, suggesting that possible biases due to the addition of a small sub-sample of low-mass clusters from the literature (i.e. the radio brightest ones) are likely marginal. 
We stress however that only future studies of statistical samples of galaxy clusters covering a larger range of cluster masses, down to few $10^{14} M_{\sun}$, have the potential to unambiguously determine the slope of the correlation.

Depending on the regression method used, the slope of the correlation ranges from 2.7 to 3.5. This is consistent with the findings of \citet{cassano13} and \citet{martinezaviles16}. As a further check, we derived the correlation parameters using the Bayesian regression method linear regression in astronomy \citep[LIRA][]{sereno16}. By default, LIRA treats X as the independent variable and Y as the dependent one. We applied LIRA to the extended sample, excluding the small radio halos and USSRH. We obtained a slope B = $2.85\pm 1.3$, which is in fact consistent with the BCES Y|X estimation. The larger uncertainty in the slope estimate with LIRA is due to the fact that this method considers a larger 
number of parameters with respect to BCES \citep{sereno16}.

\subsection{Scatter of the correlation and clusters' dynamics}
\label{sec: scatter}

\begin{figure}
\centering
\includegraphics[width=8cm]{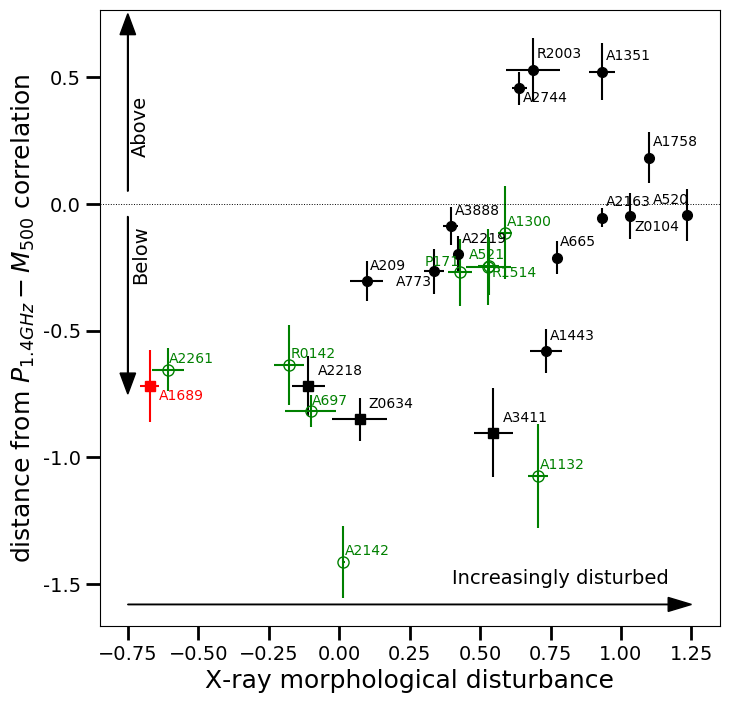}
\caption{Distance of radio halos from the correlation (BCES Y|X method) vs. X-ray morphological disturbance ( see Section \ref{sec: scatter} for details). Green empty circles represent USSRH or candidate USSRH, black squares are clusters with $R_H/R_{500}\sim 0.2$ and the red square is A1689, which also has $R_H/R_{500}\sim 0.2$ and it is ongoing a merger along the line of sight \citep{andersson04}.}
\label{Fig:distance}
\end{figure}

For all fitting methods, the correlation shows a fairly large scatter (Table \ref{Tab:fit}), slightly larger than the one found by \citet{cassano13}.
Different mergers may generate different spectra and a large range of radio emitting volumes (extension of the turbulent regions), as we also see in the right panel of Fig. \ref{Fig:histo_rh}.
The time evolution of radio halos may also contribute to the scatter of the correlation. Indeed, simulations show that the synchrotron emission evolves with the cluster dynamics, increasing in the early stage of the merger, when turbulence accelerates electrons, and then decreasing along with the dissipation of turbulence at later merger stages \citep{donnert13}. In the radio power--mass plane, this induces a migration of clusters from the region of the upper limits to the correlation (or above) and then a progressive dimming of the radio power together with a steepening of the spectrum. \citet{yuan15} suggested that the scatter of the radio power -- X-ray luminosity correlation can be significantly reduced if the dynamical state of clusters is taken into account. Focusing on the statistical sample, without the additional clusters in Table \ref{Tab:additional}, we show in Fig. \ref{Fig:distance} the distance (on the $P_{1.4 \rm{GHz}}$ axis) of radio halos from the correlation vs their X-ray morphological disturbance, measured as the distance from the bisector (the one with positive angular coefficient) of the two dashed lines shown in the $c-w$ diagram (Fig. \ref{Fig:morphological_diagrams}, panel a).
We found a clear trend between these two quantities, with radio halos scattered up with respect to the correlation being hosted in more dynamically disturbed clusters. We ran a Spearman test \citep{spearman} and obtained $r_s=0.6$ and a probability of no correlation of $1.4\times10^{-3}$. Fig. \ref{Fig:distance} suggests that the merger activity has a key role in determining the position of radio halos with respect to the correlation, thus inducing at least part of the large scatter that we observe around the correlation. This is in line with previous findings based on simulations \citep{donnert13}.

It is interesting to note that the fraction of USSRH or candidate USSRH among the most disturbed clusters in Fig. \ref{Fig:distance} is $\sim 20\%$, whereas it is $\sim 70\%$ in the less disturbed systems. Although the available statistics on USSRH is still poor and the information on the spectra is currently not homogeneous, this hint can be interpreted as the consequence of the fact that less energetic mergers induce a low level of turbulence in the ICM, which in turn is not able to accelerate particles up to the energy necessary to emit at $\sim$GHz frequencies.
These less disturbed systems can be either minor mergers or systems in their very late or very early merger state, which are expected to appear more relaxed in the X-rays and to develop steep radio spectra \citep[e.g.][]{donnert13}.
Interestingly, our results are in line with recent findings suggesting that steep-spectrum halos reside in clusters with high X-ray luminosity relative to that expected from the mass-X-ray luminosity scaling relations, indicating that such systems may be in an earlier state of the merger \citep{birzan19}. Finally, we also note that small radio halos are predominantly found in less disturbed systems, possibly suggesting that minor mergers dissipate turbulence in smaller volumes.

\subsection{Radio bimodality}
\label{Sec:bimodality}

\begin{figure}
\centering
\includegraphics[width=\columnwidth]{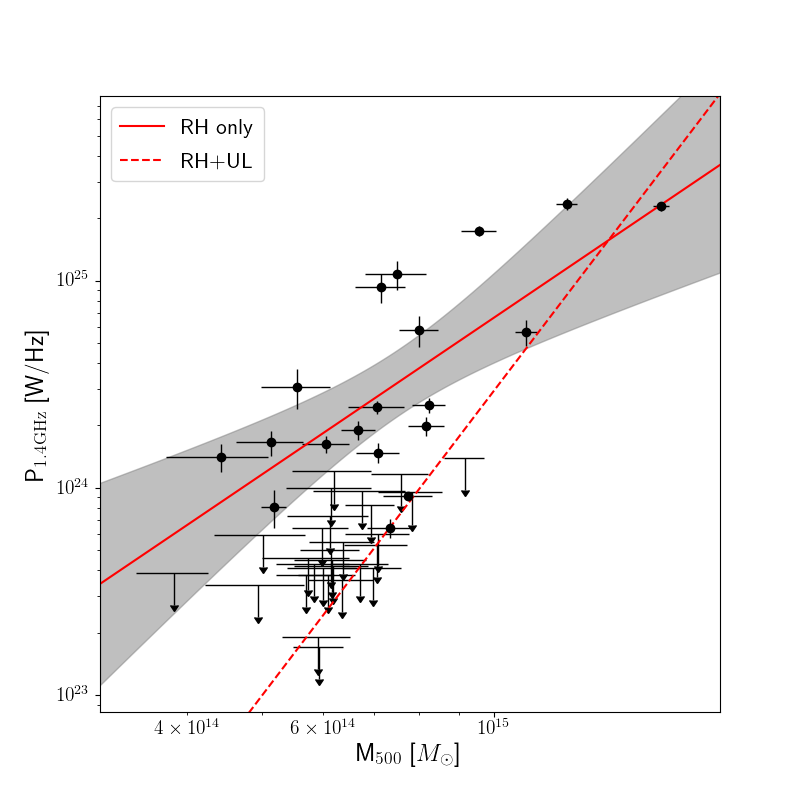}
\caption{Distribution of radio halos and upper limits in the $P_{1.4 \rm{GHz}}-M_{500}$ diagram after removing the five small radio halos. The best fit relation including only radio halos derived with the EM regression method is shown with its 95\% confidence region (red solid line and shadowed region). The dashed line is the EM best-fit relation to radio halos plus upper limits.}
\label{Fig:asurv}
\end{figure}

It is well known that merging clusters with radio halos lie on the radio luminosity--mass (or X-ray luminosity) correlation, while relaxed clusters typically do not host radio halos and their radio power lies well below the correlation \citep{brunetti07, brunetti09, cassano13}. 
Here we investigate the presence of a bimodal behaviour of clusters with and without radio halos in the radio power-mass diagram focusing on the extended sample without small radio halos (Fig. \ref{Fig:P_M_Cuciti}, right). We found that, for clusters with $M_{500}\gtrsim5.5\times10^{14}M_\odot$, almost all the upper limits ($\sim90\%$) are below the 95\% confidence region of the correlation, confirming previous studies \citep[e.g.][]{cassano13}.

To further investigate the radio distribution of clusters with and without radio halos in the $P_{1.4 \rm{GHz}}-M_{500}$ diagram, we made use of another method of regression analysis, based on the Expectation Maximization (EM) algorithm, that is implemented in the ASURV package \citep{isobe86} and deals with upper limits as `censored data'. We applied the EM regression algorithm to the radio halos only and then to the combined radio halos and upper limits. The resulting best fits are shown in Fig. \ref{Fig:asurv} and are summarised in the bottom lines of Table \ref{Tab:fit}. The correlation obtained including upper limits in the fit (dashed line) is much steeper than the one derived using only detections. This is due to the fact that the EM algorithm allows for the `censored data' to assume values smaller than those of the upper limits in an iterative process whose aim is to find a maximum likelihood solution. This, in addition to the fact that upper limits dominate the cluster statistics at low masses and that the bulk of them is clustered quite below the detections, leads to a best-fit that is significantly different from the bet-fit line to radio halos. 
This hints at the existence of two distinct radio state of galaxy clusters: `on-state' radio halo clusters that lie on the correlation, and `off-state' clusters that occupy a different region of the diagram \citep[see also][]{brown11}.

Clusters with upper limits at GMRT frequencies have been extrapolated to 1.4 GHz with $\alpha=-1.3$. Here we point out that a flatter spectrum ($\alpha\sim-1.1$) would make limits only $\sim 20\%$ shallower with no impact on the conclusions of this Section. On the other hand, a steeper spectrum would make limits even deeper, increasing the separation from radio halos.

\subsection{Emissivity of radio halos}
\label{Sec:emissivity}

\begin{table}
\begin{center}
\begin{small}
\caption{Emissivity of radio halos and upper limits}
\begin{tabular}{lc}
\\
\hline
name &  $J$ \\
     & ($10^{-42}$erg s$^{-1}$ cm$^{-3}$ Hz$^{-1}$)\\
\hline\\
A209 & $ 0.44 \pm 0.04 $ \\
A665 & $ 0.67 \pm 0.07 $ \\
A773 & $ 0.37 \pm 0.05 $ \\
A2163 & $ 0.92 \pm 0.05 $ \\
A2218 & $ 2.36 \pm 0.32 $ \\
A2744 & $ 2.82 \pm 0.22 $ \\
A2219 & $ 0.99 \pm 0.09 $ \\
Z0634 & $ 0.88 \pm 0.16 $ \\
A1758 & $ 1.99 \pm 0.25 $ \\
A2345 & $ <0.28 $ \\
A2104 & $ <0.11 $ \\
R0616 & $ <0.11 $ \\
A2895 & $ <0.21 $ \\
A56 & $ <0.54 $ \\
A2355 & $ <0.36 $ \\
A1733 & $ <0.23 $ \\
PSZG019 & $ <0.27 $ \\
A2813 & $ <0.61 $ \\
A384 & $ <0.31 $ \\
\hline\\
A520 & $ 1.94 \pm0.14 $ \\
Z0104 & $ 1.11 \pm0.10 $ \\
A1451 & $ 0.88 \pm0.09 $ \\
A3888 & $ 1.59 \pm0.16 $ \\
A3411 & $ 1.59 \pm0.54 $ \\
A1689 & $ 2.56 \pm0.69 $ \\
A1443 & $ 0.64 \pm0.02 $ \\
A1576 & $ <0.30 $ \\
A2697 & $ <0.19 $ \\
R0142 & $ <0.21 $ \\
A1423 & $ <0.18 $ \\
A2537 & $ <0.21 $ \\
A68 & $ <0.20 $ \\
A781 & $ <0.17 $ \\
A3088 & $ <0.20 $ \\
A2631 & $ <0.19 $ \\
  
\hline
\label{Tab:emissivity}
\end{tabular}
\begin{flushleft}
Notes -- Top panel: clusters with available fitting parameters $I_0$ and $r_e$; Bottom panel: remaining clusters, for which $I_0$ and $r_e$ are estimated from the integrated flux within $R_H$.
\end{flushleft}
\end{small}
\end{center}
\end{table}

To study the distribution of the clusters in the radio power mass diagram we removed
the smallest radio halos, whose volume is $6-20$ times smaller than the average volume of radio halos in the sample. In general,
the different emitting volumes of radio halos may drive a significant fraction of the
scatter in the $P_{1.4 \rm{GHz}}-M_{500}$ diagram. Therefore, one possibility to remove this
effect is to look at the emissivity, instead of the radio luminosity. In the following, we will focus on the statistical sample, without the additional clusters in Table \ref{Tab:additional}.

In paper I, we derived the azimuthally averaged surface brightness radial profile of radio halos and we fitted them with an exponential law to obtain the central surface brightness and the $e$-folding radius, $r_e$. Following \citet{murgia09}, for these radio halos we calculated the volume averaged emissivity by assuming that their flux density (obtained by integrating the best fit exponential profile up to $3r_e$) comes from a sphere of radius $3r_e$:
\begin{equation}
\label{eq:emissivity}
    J\simeq 7.7\times 10^{-41} (1+z)^{3+\alpha}~ \frac{I_0}{r_e}~~~~~~  [\mathrm{erg} ~\mathrm{s}^{-1} \mathrm{cm}^{-3} \mathrm{Hz}^{-1}]
\end{equation}
In the same way, we calculated the upper limits to the emissivity of clusters without radio halos. In particular, we could use this approach for the upper limits derived in paper I by injecting an exponential model into the data because we choose $r_e=500/2.6=192$ kpc\footnote{The value 2.6 was derived in \citet{bonafede17} to convert $R_H$ into $r_e$.} and $I_0$ is the one corresponding to the upper limit flux.
The values obtained for the emissivity are reported in Table \ref{Tab:emissivity} (top panel).

\begin{figure}
\centering
\includegraphics[width=8cm]{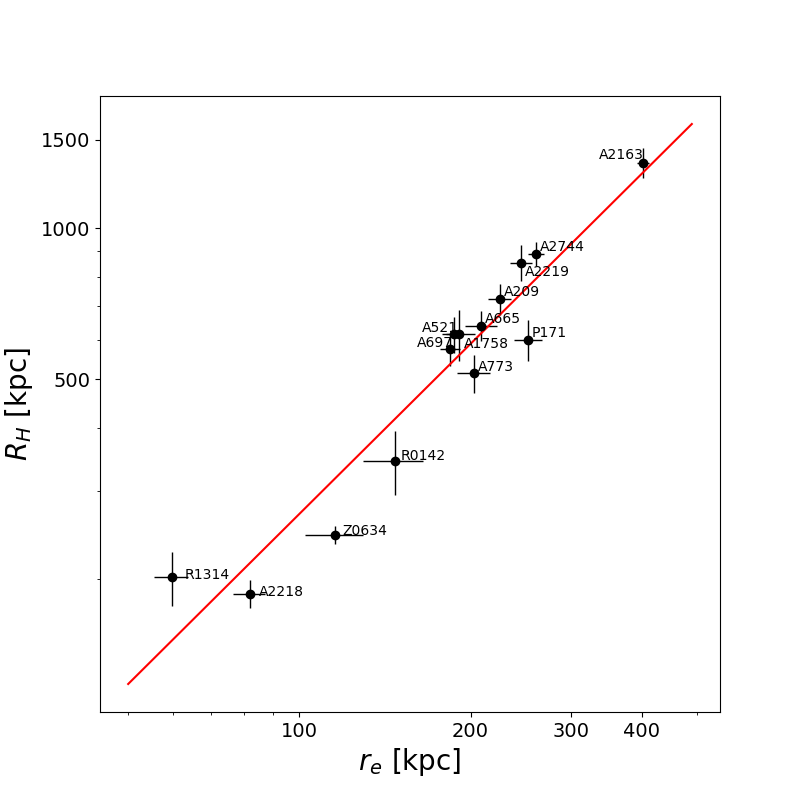}
\caption{Radio halos radius measured as in Section \ref{sec:fitting_results} ($R_H$) vs. $r_e$ of the clusters with available surface brightness radial profile. We show the best fit line in red.}
\label{Fig:rerh}
\end{figure}

\begin{figure*}
\centering
\includegraphics[width=\textwidth]{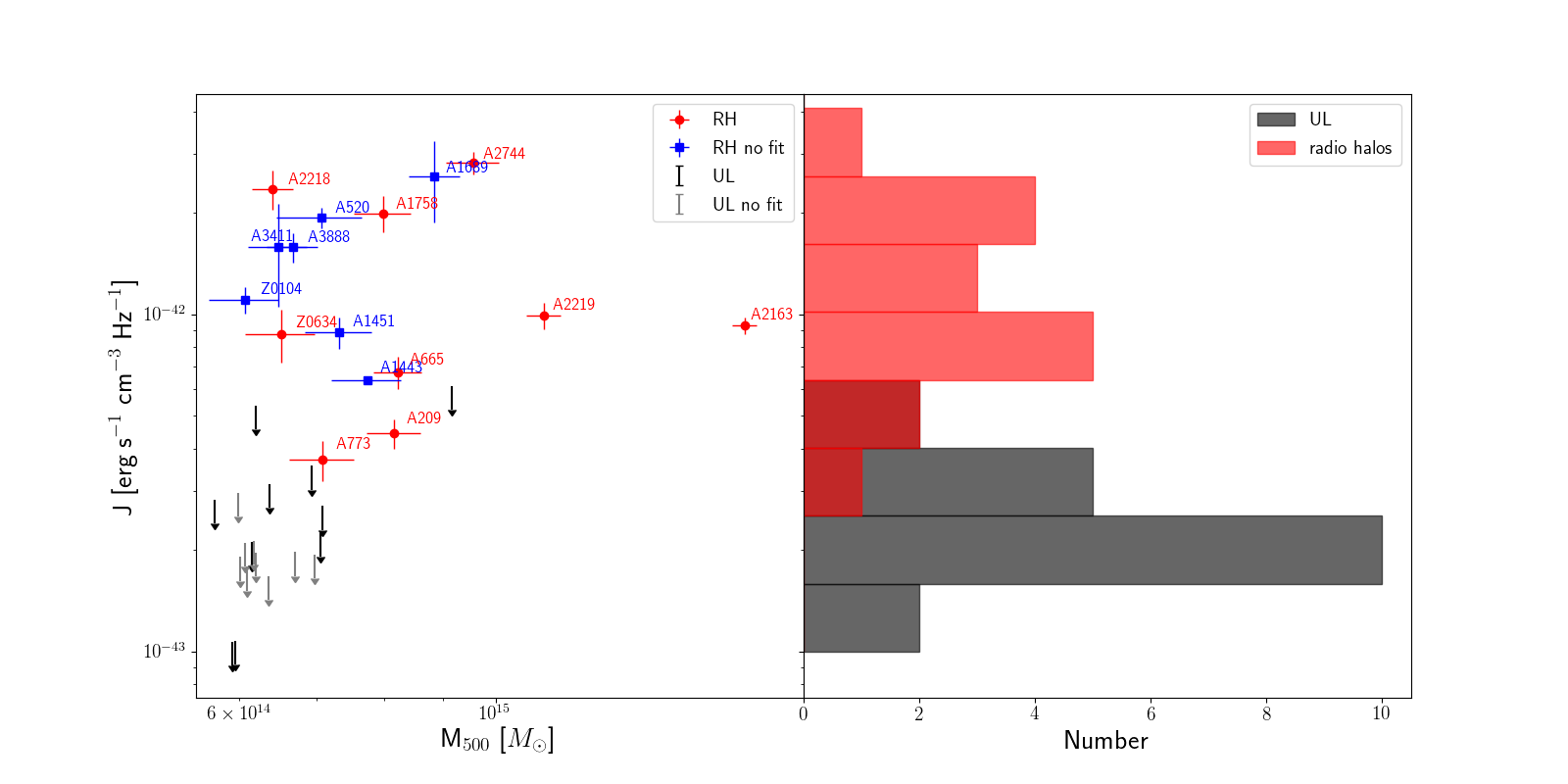}
\caption{Emissivity vs. mass diagram (left) and emissivity distribution (right). Red circles are radio halos with available radial profile while blue squares are the remaining radio halos. Upper limits derived in paper I injecting an exponential model are represented with black arrows, while upper limits from the literature are the gray arrows.}
\label{Fig:emissivity}
\end{figure*}

We were able to derive the emissivity with this approach only for radio halos with a single peak and a regularly decreasing brightness (about half of the radio halos in our sample). Here we estimate the emissivity also for the remaining radio halos, assuming that the measured flux comes from a sphere of radius $R_H$, where $R_H$ is the one derived in Section \ref{sec:fitting_results}. 
In Fig. \ref{Fig:rerh}, we show $R_H$, as derived in Section \ref{sec:fitting_results}, as a function of $r_e$ for the radio halos that have a radial profile available. $R_H$ appears to be well correlated with $r_e$, in particular, a simple least squares fit gives log$(R_H)=1.12\times \mathrm{log}(r_e)+0.18$. We used this relation to infer $r_e$ for the remaining radio halos and we estimated $I_0$ by assuming that the measured flux corresponds to the integral of an exponential function up to $R_H$. To estimate the emissivity of the upper limits taken from the literature, we assumed $r_e=192$ kpc, to be consistent with the upper limits derived in paper I. We note that if we use the correlation shown in Fig. \ref{Fig:rerh} to estimate $r_e$ for the literature upper limits, assuming $R_H=500$ kpc, we would obtain an offset in the emissivity of less than 10\%. We report the values of the emissivity for the remaining radio halos and upper limits in the bottom panel of Table \ref{Tab:emissivity}.

We show the distribution of radio halos and upper limits in the emissivity-mass diagram in Fig. \ref{Fig:emissivity}. In this plot we do not show the 4 upper limits at $z>0.31$ because, although they are similar to the rest of the upper limits in terms of flux, their luminosity and emissivity is significantly higher (equation \ref{eq:emissivity}), meaning that the sensitivity of our current observations does not allow to put stringent upper limits at high redshift. For consistency, we removed also the two halos at $z>0.31$ (A1351 and R2003). This is the first time that a systematic study of the emissivities of radio halos is performed in a statistical sample. We found that upper limits have emissivity $\sim10$ time smaller than the bulk of radio halos. The apparent bimodality shown in the right panel of Fig. \ref{Fig:emissivity} is currently driven by the sensitivity of our observations (which sets the level of our upper limits) and can reflect two possible intrinsic distributions: {\it i)} the emissivity of clusters without radio halos is close to the upper limit value, meaning that the distribution is truly bimodal and future more sensitive observations should be able to detect these radio halos, {\it ii)} the emissivity of clusters without radio halos is for the most part much smaller than the upper limits, in this case the upper limits would represent a tail of the radio halos emissivities extending down to 1-2 orders of magnitude below and cluster would be `lifted up' from the tail to the bulk of radio halos as a consequence of merger events. Looking at this point with deeper observations in the future we will be able to provide valuable information on the evolution of halos and on the level of the hadronic contribution \citep[see][]{cassano12}.

\section{Occurrence of radio halos}
\label{Sec:occurrence}


To measure the occurrence of radio halos as a function of the cluster mass, we focus on the statistical sample. The great majority of the galaxy clusters with radio upper limits are below the 95 percent confidence level spanned by the radio power-mass correlation (Fig. \ref{Fig:P_M_Cuciti}, right); we classify these clusters as non-radio halos. In five cases (A1914, A115, PSZ1 G205.07-6294, A1763 and A2390), clusters are classified as non-radio halo from the literature using deep radio observations; although no upper limits were derived, we classified also these cases as non-radio halos.
However, there are two upper limits consistent with the correlation in our sample, and four clusters for which we were not able to derive upper limits due to problems in the observations. Moreover there are five candidate radio halos in the sample. These 11 clusters have uncertain classification. We will consider two cases: a `reduced sample', which does not include the uncertain cases at all and the total sample, where we assumed that candidate radio halos actually host radio halos and clusters without good upper limits do not host radio halos. We point out that a random assignment of radio halos among the 11 uncertain clusters would lead to almost to the same measured fractions of halos. 

Following the procedure adopted in \citet{cuciti15}, we split the sample into two mass bins and measured the fraction of clusters with radio halos, $f_{RH}$, in the low-mass bin (LM, $M<M_{lim}$) and in the high-mass bin (HM, $M>M_{lim}$). In particular, we used $M_{lim}=8\times10^{14}M_\odot$ for consistency with our previous study. With this partition we have:
\begin{itemize}
\item[-] 60 clusters in the LM bin, among which 9 host radio halos, 5 host USSRH or candidate USSRH, 3 host radio halos with $R_H/R_{500}\sim$0.2 and 11 have uncertain classification (5 candidate radio halos and 6 clusters without solid upper limit).
\item[-] 15 clusters in the HM bin, 5 of which have radio halos, 5 have USSRH or candidate USSRH, one has a radio halo with $R_H/R_{500}\sim$0.2.
\end{itemize}

In total, there are four small radio halos in the statistical sample, one of the small halos shown in Fig. \ref{Fig:histo_rh} belongs to the extended sample. 
In order to take into account the uncertainty on the derived fractions of radio halos associated to the statistical error on the masses, we used a Monte Carlo approach. Specifically, we randomly extracted the mass of each cluster from a Gaussian distribution having median value $\mu=M_{500}$ and standard deviation $\sigma=\sigma_{M_{500}}$ where $M_{500}$ and $\sigma_{M_{500}}$ are the values of the mass and associated error as reported in the Planck catalog\footnote{since the errors on the masses are not symmetric, here we assumed $\sigma_{M_{500}}$ to be equal to the largest error.}. Then, we split the clusters into the two bins with $M_{lim}=8\times10^{14}M_\odot$ and we calculated the fractions of radio halos. We repeated this procedure 1000 times and we assumed that the fraction of radio halos in each bin is the mean of the resulting distribution and the error is the standard deviation. With this approach we obtained that the fraction of clusters with radio halos in the LM bin is $f_{RH}=37\pm2$\% in the total sample and $f_{RH}=35\pm2$\% in the `reduced sample', while in the HM bin $f_{RH}=67\pm6$\% both in the total and reduced samples. We thus confirmed the existence of a drop in $f_{RH}$ at low mass systems with a complete ($\gtrsim80$\% mass completeness) mass-selected sample of galaxy clusters.


\subsection{Comparison with theoretical expectations}
\label{Sec:model}

\begin{figure}
\centering
\includegraphics[width=\columnwidth]{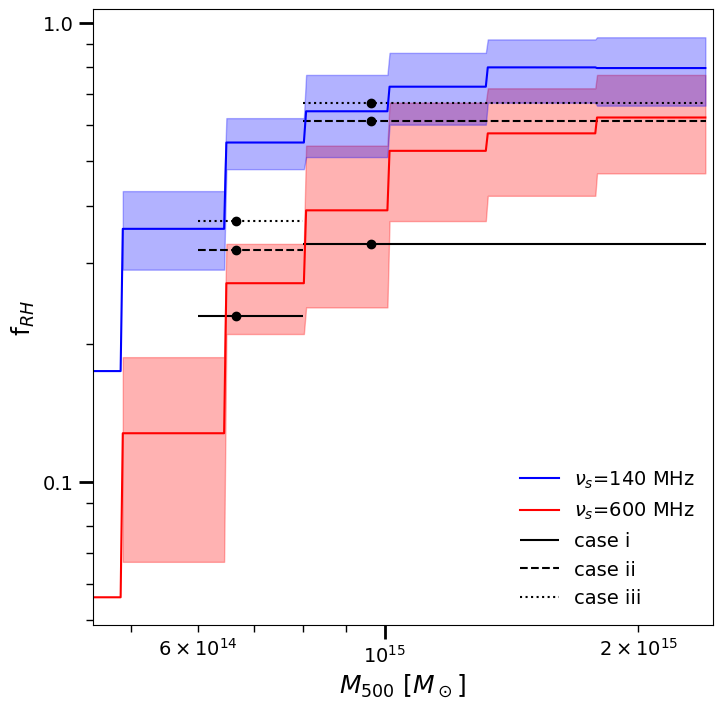}
\caption{\small {Expected fraction of clusters with radio halos with steepening frequency $\nu_s>600$ MHz and $\nu_s>140$ MHz in the redshift range $0.08<z<0.33$ (red and blue lines, respectively). Shadowed regions represent the uncertainty on the model predictions taking into account the statistical error associated to the limited size of the observed sample. Calculations have been performed for the following choice of model parameters: $b=1.5$, $\langle B\rangle=1.9~\mu$G (where $B=\langle B\rangle\times (M/\langle M\rangle)^b),$ and the fraction of energy channelled into particle acceleration $\eta_t=0.2$ \citep[see][and references therein]{cassano12}. The observed fraction of clusters with radio halos in the two mass bins is overlaid (black lines) and refers to the three cases in Tab. \ref{Tab:fraction} for the total sample. The dots represent the average mass of the clusters in the bins.}}
\label{Fig:frh}
\end{figure}

We used the model developed by \citet{cassanobrunetti05} and \citet{CBS06} to derive the formation probability of radio halos as a function of the mass of the host cluster in the redshift range of our sample ($z=0.08-0.33$). These models take into account the formation history of galaxy clusters using merger trees and calculate the generation of turbulence, the particle acceleration and the synchrotron spectrum during the clusters' lifetime. These models currently offer the unique possibility to calculate the expected occurrence of radio halos to be compared with observations.
The basic idea of these models is that the synchrotron spectra of radio halos are characterised by a steepening frequency, $\nu_s$, which is the result of the competition between turbulent acceleration and radiative (synchrotron and Inverse Compton) losses. 
In general, $\nu_s\gtrsim$ GHz is expected in the most massive clusters, undergoing major mergers, while less energetic merger events, involving clusters with smaller masses, are expected to form radio halos with lower values of $\nu_s$. These radio halos, should show extremely steep spectra ($\alpha<-1.5$, with $S(\nu)\propto \nu^{\alpha}$) when observed at $\sim$GHz frequencies and they are expected to constitute the class of USSRH. The possibility to detect a radio halo is thus related to the observing frequency, $\nu_{obs}$, in particular, the spectral steepening challenges the detection of radio halos with $\nu_s <\nu_{obs}$.



We calculated the theoretical formation probability of radio halos with $\nu_s>$ 600 MHz and $\nu_s>$ 140 MHz as a function of the cluster mass in the redshift range $z=0.08-0.33$ (Fig. \ref{Fig:frh}). The uncertainties in the model (red and blue shadowed regions) are calculated with Monte Carlo extractions from the large number of theoretical merger trees and take into account the statistical error introduced by the limited size of our observed sample. The value of $\nu_s>$ 600 MHz can be considered as a reference frequency for both VLA 1.4 GHz and GMRT 610 MHz observations, that constitute the great majority of the available radio observations for the clusters of our sample. 

In order to properly compare these expectations with our observations, we need to take two main points into account: 1) some of the USSRH and candidate USSRH observed in our sample may have $\nu_s<$ 600 MHz, in that case, they should be `counted' as radio halos only in the comparison with model expectations with $\nu_s>$140 MHz. 2) A limitation of these models is that the size of the emitting volume is fixed at $R_H=500$ kpc\footnote{$R_H=500$ kpc corresponds to $R_H/R_{500}\sim 0.4$ for the typical masses of our cluster sample, which is consistent with the average/typical sizes of radio halos in our sample (Fig. \ref{Fig:histo_rh}, right)}. However, the volume of the small radio halos in our sample is $6-20$ times smaller than the volume assumed in models, implying that the occurrence of radio halos in these situations is biased low in models, as a result of the fact that turbulence generated in a smaller volume is artificially spread in a Mpc$^3$ causing a decline of particle acceleration efficiency and $\nu_s$. Therefore, in the comparison between models and observations, one option is to consider small radio halos as non radio halos or, in alternative, as USSRH. 

For these reasons, we consider three possibilities for the comparison with calculations that assume a value of the minimum steepening frequency $\nu_s$: i) small radio halos and USSRH are non radio clusters (i.e. their steepening frequency is assumed to be lower than the minimum steepening frequency in the considered models), ii) small radio halos are non radio halo clusters whereas USSRH are, iii) both small radio halos and USSRH are considered as radio halo clusters. To simplify the comparison, here we consider candidate USSRH as USSRH, although future observations might not confirm their steep spectra. In Table \ref{Tab:fraction} we report the observed fractions of radio halos in these three cases both for the reduced and the total sample. These fractions are derived with the Monte Carlo approach described in Section \ref{Sec:occurrence}. We show the comparison between the predicted and observed fraction of radio halos as a function of mass in Fig. \ref{Fig:frh}. The difference between the total and the `reduced' samples in terms of $f_{RH}$ is marginal (Table \ref{Tab:fraction}), thus, in Fig. \ref{Fig:frh}, we report only the total sample, for clarity.

\begin{table*}
\begin{footnotesize}
\begin{center}
\caption{Observed fraction of radio halos}
\begin{tabular}{|l|cc|cc|cc|}
\hline
&&&&&&\\
	&	\multicolumn{2}{c|}{i}	&	\multicolumn{2}{c|}{ii}	&	\multicolumn{2}{c|}{iii}\\[2ex]
\hline
&&&&&&\\
	&		$f_{RH}$(LM)	&	$f_{RH}$(HM)	&	$f_{RH}$(LM)	&	$f_{RH}$(HM)	&	$f_{RH}$(LM)	&	$f_{RH}$(HM)\\[2ex]
\hline
&&&&&&\\
total sample&	$0.23\pm0.02$&	$0.33\pm 0.05$&	$0.32\pm0.02$&	$0.61\pm0.06$&  $0.37\pm0.02$	&	$0.67\pm0.06$\\[3ex]
reduced sample 	&	$0.18\pm 0.02$	&	$0.33\pm 0.05$&	$0.29\pm 0.02$&	$0.61\pm0.06$& $0.35\pm0.02$	&	$0.67\pm0.06$\\
&&&&&&\\
\hline
\end{tabular}	
\label{Tab:fraction}	
\end{center}
\tablefoot {LM = Low Mass bin ($M_{500}<8\times10^{14}M_\odot$), HM = High Mass bin ($M_{500}>8\times10^{14}M_\odot$). i: small radio halos and USSRH are considered as non radio halo clusters; ii: small radio halos are considered as non radio halo clusters, USSRH are considered as radio halo clusters; iii: small radio halos and USSRH are considered as radio halo clusters.}
\end{footnotesize}
\end{table*}

In spite of the basic assumption adopted in the model there is a remarkable agreement between the observed fraction of clusters with radio halos in the two mass bins and the model expectation with $\nu_s>$ 600 MHz. In particular, case i) can be considered the lower limit inferred from observations in the comparison with this particular model, essentially because a fraction of the USSRH and candidate USSRH in our sample might have $\nu_s > 600$ MHz. Conversely, case ii) can be considered as the upper limit in this comparison. On the other hand, when attempting a comparison with models with $\nu_s > 140$ MHz (blue line) cases ii) and iii) provide the most relevant observational constraints. These constraints are however driven by observations at higher frequencies (600-1400 MHz, paper i) and may have lost a significant number of USSRH in our sample.
As a consequence, the discrepancy between the model predictions for $\nu_s> 140$ MHz and the observed fraction of radio halos in the LM bin implies that a number of radio halos in low mass clusters should be discovered with future low frequency observations.  


These models are anchored to the observed mass distribution function of clusters \citep{press74} and to the rate of mergers that is predicted by the standard $\Lambda$CDM model and consequently the fact that at low mass, the occurrence of radio halo decrease, implies that the number of merging clusters without radio halos (when observed at frequencies $> 600$ MHz) should increase. In agreement with that, in Section \ref{Sec:connection} we found that the fraction of merging clusters without radio halos is much higher ($\sim65\%$) in low mass clusters with respect to high mass ones ($\sim20\%$). At the same time, this model predicts that a fraction of these mergers will generate radio halos emitting at lower frequencies \citep[e.g.][see also Fig. \ref{Fig:frh}]{CBS06,cassano10a,cassano12}. Consequently, we expect that a large fraction of the merging systems in our sample that do not show radio halos in the GMRT and JVLA images will be USSRH that can be detected by low frequency observations. In this respect, 
with the LOw Frequency Array \citep[LOFAR,][]{vanhaarlem13} we will perform the statistical analysis of radio halos at low frequencies and the comparison with the results presented in this paper will give unprecedented insight on the formation and evolution of radio emission in clusters.

\section{Summary and Conclusions}
\label{Sec:conclusion}

We presented the first statistical analysis of radio halos in a mass-selected sample of 75 galaxy clusters. Clusters were selected from the Planck SZ catalogue \citep{planck14} with mass $M_{500}\geq6\times10^{14} M_\odot$ and redshift $z=0.08-0.33$. The radio and X-ray data analysis of these clusters are described in paper I. In the following, we summarise the main steps and results of the statistical analysis performed in this paper.

\begin{itemize}
\item We combined the radio information on the clusters of our sample with the dynamical information coming from the X-ray data analysis (see paper I) in the morphological diagrams (Fig. \ref{Fig:morphological_diagrams}). We found that more than the 90\% of radio halos are in merging galaxy clusters, whereas less than 10\% is in non merging systems. As expected, not all the merging clusters host a radio halo. Interestingly, the fraction of merging clusters without radio halos is $\sim65$\% in low mass clusters, whereas it is $\sim20$\% in high mass ones (Fig. \ref{Fig:c_w_massa}). This is in line with turbulent re-acceleration models predicting that merging events in low mass systems may not induce enough turbulence to accelerate particles emitting at $\sim$GHz frequencies.

\item We confirmed the presence of a correlation between the radio power of radio halos and the mass of the host clusters (Fig. \ref{Fig:P_M_Cuciti}). However, we showed that the small range of masses of our sample is a significant limitation if we aim at constraining the slope of the correlation (App. \ref{app:bces}). Therefore, we considered an `extended' sample, made by the combination of our statistical sample with a subsample of clusters from \citet{cassano13} and \citet{martinezaviles16}. Depending on the fitting method, the slope of the correlation ranges between 2.7 to 3.5. This result is consistent with previous findings \citep{cassano13, martinezaviles16}. 

\item We investigated the possible connection between the scatter of the correlation and the different dynamical state of clusters.
We found a clear trend between the distance from the $P_{1.4 \rm{GHz}}-M_{500}$ correlation and the dynamical disturbance of clusters (Fig. \ref{Fig:distance}). This indicates that the large scatter around the correlation is at least partly due to the complex mix of merging histories in the diagram. Interestingly, the great majority of radio halos in less disturbed systems are USSRH (Fig. \ref{Fig:distance}), suggesting that less energetic merger events may not be sufficient to produce radio halos emitting at $\sim$ GHz frequencies. Additionally, some of these clusters may be in a late stage of merger where the X-ray morphology appears relaxed and the radio halo spectrum has steepened.  

\item Although the scatter of the correlation is relatively large and the bimodality is less evident, still limits occupy a different region of the $P_{1.4 \rm{GHz}}-M_{500}$ diagram, being below the 95\% confidence region of the correlation (Fig.\ref{Fig:P_M_Cuciti} and Fig. \ref{Fig:asurv}). 

\item For the first time, we studied the emissivity of radio halos in a mass-selected sample of clusters. In the emissivity, the scattering induced in the radio luminosities by the different emitting volume is reduced. Indeed, we find a clear separation between radio halos and upper limits with limits lying more than five times below the bulk on radio halos in the emissivity-mass diagram.

\item Following \citet{cuciti15}, we measured the occurrence of radio halos in two mass bins: $M<8\times10^{14}M_\odot$ and $M\geq8\times10^{14}M_\odot$. We found that the fraction of clusters with radio halos is $\sim 70$\% in high mass clusters and $\sim 35$\% in low mass clusters, thus confirming the presence of a drop in the fraction of radio halos in low mass systems. 

\item We used the model developed by \citet{cassanobrunetti05} to compare the observed and predicted fraction of radio halos as a function of mass (Fig. \ref{Fig:frh}). Although these models have some limitations that require a careful comparison, we showed that there is good agreement between the theoretical expectations and our observations. The combination of the decline of occurrence and increase of the fraction of merging clusters without radio halos in lower mass clusters suggests that a population of USSRH should be discovered among those systems with low frequency observations.
\end{itemize}
 
 \begin{acknowledgements}
The authors thank the anonymous referee for the useful comments which improved the presentation of the paper.
VC acknowledges support from the Alexander von Humboldt Foundation. RJvW acknowledges support from the VIDI research programme with project number 639.042.729, which is financed by the Netherlands Organisation for Scientific Research (NWO). RK acknowledges the support of the Department of Atomic Energy, Government of India, under project no. 12-R\&D-TFR-5.02-0700. Basic research in radio astronomy at the Naval Research Laboratory is supported by 6.1 Base funding. SE acknowledges financial contribution from the contracts ASI-INAF Athena 2019-27-HH.0,
``Attivit\`a di Studio per la comunit\`a scientifica di Astrofisica delle Alte Energie e Fisica Astroparticellare''
(Accordo Attuativo ASI-INAF n. 2017-14-H.0), INAF mainstream project 1.05.01.86.10, and
from the European Union’s Horizon 2020 Programme under the AHEAD2020 project (grant agreement n. 871158). GWP acknowledges the support of the French space agency, CNES.
The National Radio Astronomy Observatory is a facility of the National Science Foundation operated under cooperative agreement by Associated Universities, Inc. We thank the staff of the GMRT that made these observations possible. GMRT is run by the National Centre for Radio Astrophysics of the Tata Institute of Fundamental Research. The scientific results reported in this article are based in part on data obtained from the \textit{Chandra} Data Archive. This research has made use of the NASA/IPAC Extragalactic Database (NED) which is operated by the Jet Propulsion Laboratory, California Institute of Technology, under contract with the National Aeronautics and Space Administration.
\end{acknowledgements}

\bibliographystyle{aa} 
\bibliography{biblio_virgi} 

\clearpage
\onecolumn
\begin{appendix}
\clearpage

\section{Test BCES}
\label{app:bces}
The sample of galaxy clusters analysed in this paper is selected from the Planck SZ catalogue imposing a cut in mass at $M_{500}\geq6\times10^{14} \,M_\odot$. As a consequence, our sample contains clusters within a relatively small range of masses ($6\leq M_{500}<16\times10^{14} \,M_\odot$, with only one cluster with $M_{500}>11\times10^{14} \,M_\odot$). In order to test whether and how such a cut influences the correlation parameters derived with the BCES methods, we performed Montecarlo simulations. We randomly distributed 60, 100 or 1000 points in the X-Y diagram on a correlation in the form Y$ = \mathrm{slope}\times \mathrm{X}$. We used five values for the slope: 3.5, 4, 4.5, 5, 5.5. We worked in log space and we choose values for the X variable similar to the ones that we are dealing with in our cluster sample, but in a larger range (X is in the range $14-15.2$, with X = log($M_{500}$)). Here the assumption is that the scaling remains a power law also extending to lower masses. We associated a random error to these points, both in X and Y, both in line with the errors that we have in our mass and radio power measurements. Then, we distributed the points around the correlation following two approaches: 
\begin{itemize}
\item[a)] we assumed that the scatter in X is only statistical scatter, while the scatter in Y is both statistical and intrinsic. 
\item[b)] we assumed that the raw scatter (statistic and intrinsic) is orthogonal with respect to the correlation line.
\end{itemize}

In both cases, the value of the scatter that we introduced was chosen to be similar to the one that we observe in our sample.
We repeated the random distribution 500 times for each value of the slope and each time we performed the BCES analysis as described is Section \ref{sec:fitting_procedure}. We found that all the BCES methods are able to recover the true slope of the correlation with $<10$\% discrepancies (the discrepancy decreases as the number of points increases).

\begin{figure}[ht]
\centering
\includegraphics[scale=0.38]{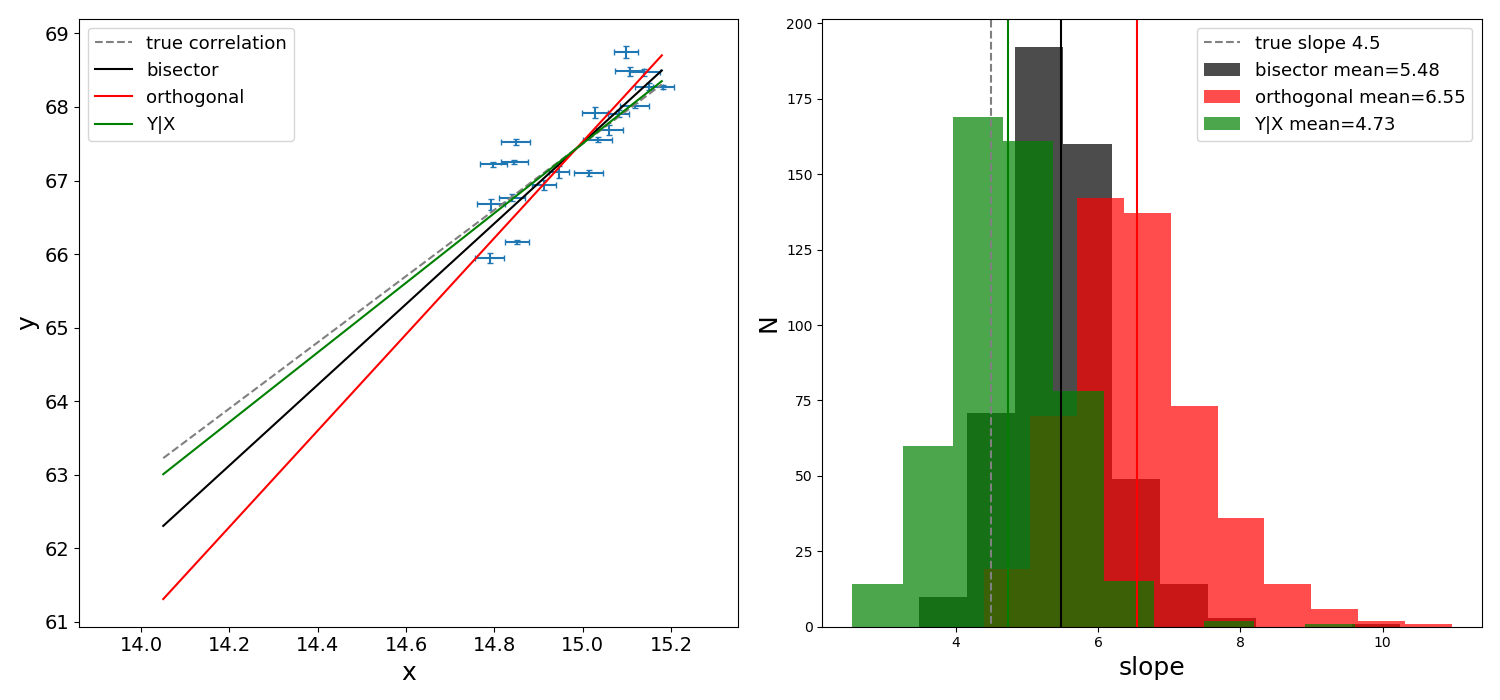}
\includegraphics[scale=0.32]{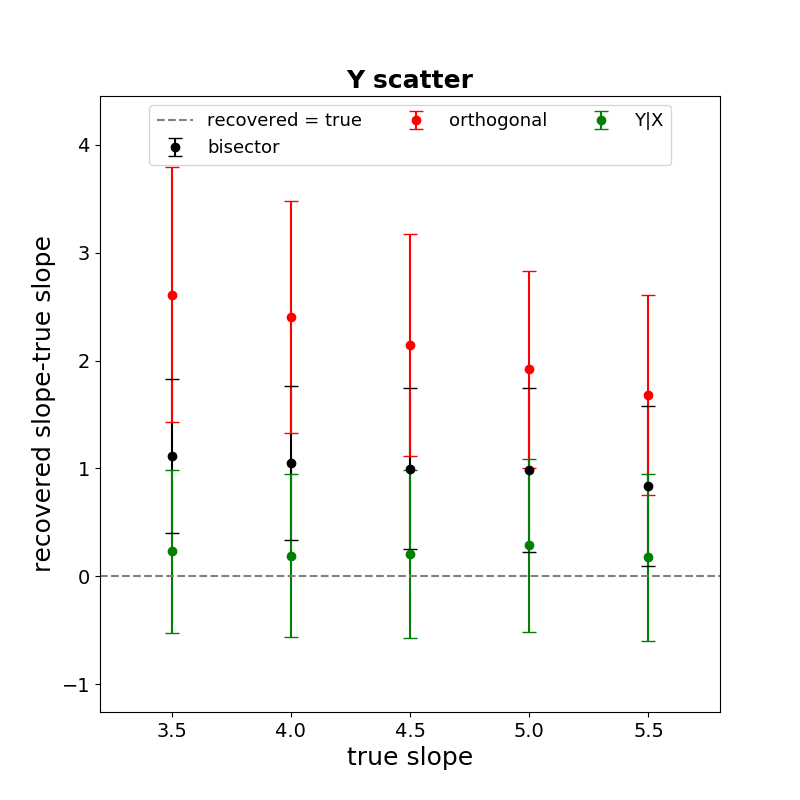}
\hspace{1cm}
\includegraphics[scale=0.32]{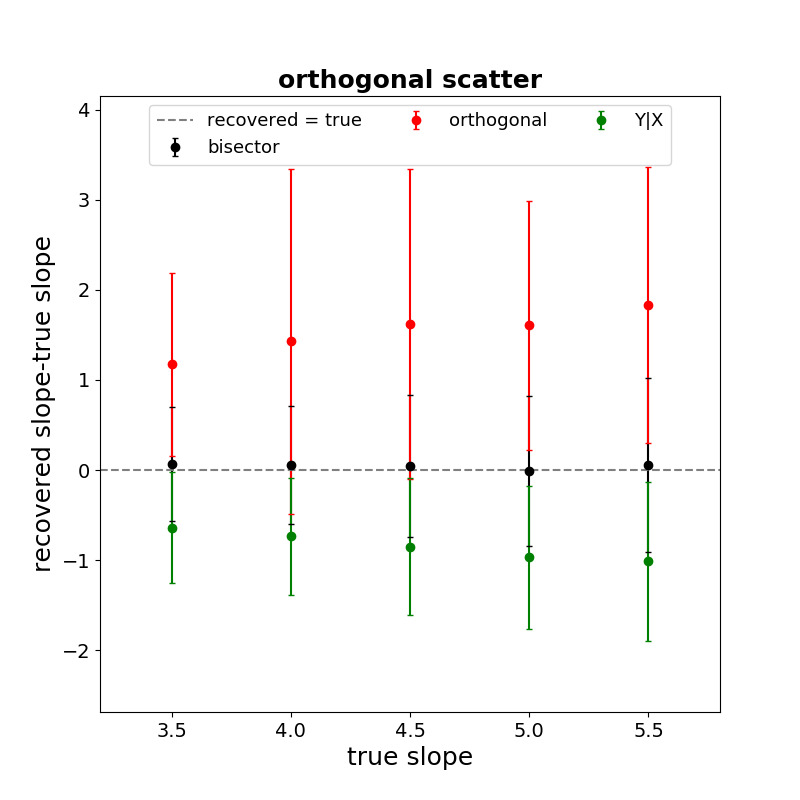}
   
\caption{{\it Top panel}: On the left, one example of the 500 Montecarlo runs with true correlation slope 4.5 and X cut at 14.77 (corresponding to a mass cut of $6\times 10^{14}$). On the right, distribution of the recovered slopes with the bisector (black), orthogonal (red) and Y|X (green) methods. Vertical lines represent the mean value of the recovered slopes and the dashed line is the true slope. These mean values of the slopes are used to draw the correlation lines in left panel. {\it Bottom panels}: Difference between recovered and true slope as a function of the true slope for the three fitting methods. The dots represent the mean values of the distributions, while the errorbars are as large as the standard deviation of the distribution of the recovered slopes. The horizontal gray line represents the case where the recovered slope matches the true slope. On the left panel we used the Y scatter, on the right the orthogonal scatter.} \label{Fig:test_bces}
\end{figure}

As a second test, after applying the scattering, we selected only points with X $\geq14.77$, corresponding to the cut in mass of our sample. An example of the distribution of points in the X-Y diagram after the cut in the X variable is shown in Fig. \ref{Fig:test_bces}, top left. In this case the true slope of the correlation is 4.5 and the scatter follows approach a). The initial distribution is made of 60 clusters, this allows us to obtain a similar number of points as in our sample, once the cut is applied. The top right panel of Fig. \ref{Fig:test_bces} shows the distribution of the slopes given by the three different fitting methods for the 500 Montecarlo trials. In Fig. \ref{Fig:test_bces}, bottom left, we show the difference between the recovered and the true slope as a function of the true slope. We repeated this test using an orthogonal scatter (approach b). Results are shown in Fig. \ref{Fig:test_bces}, bottom right panel. 
Our analysis suggests that a cut in the X variable influences the results of the fitting using the BCES algorithms. In particular, regardless of how the points are scattered around the correlation, the orthogonal method always gives steeper slopes with the largest uncertainties. Which fitting method performs better depends on how the scatter is implemented. Since it is not trivial to know what is the best way to describe the scatter around the correlation, in this paper we report the values obtained with the Y|X and the bisector methods. Also, we point out that the bisector method allows us to better compare our results with previous works \citep[e.g.][]{cassano13}.  

\end{appendix}

\end{document}